\documentclass[10pt,bibnotes,notitlepage,final,balancelastpage,pra,superscriptaddress,twocolumn,showpacs]{revtex4}
\usepackage{graphicx}
\usepackage{amssymb, amsmath, amsfonts, color, rotating, multirow, graphicx, bm}
\usepackage[dvipdfm,bookmarks=false,pdfstartview=FitH,hyperindex=true, colorlinks, linkcolor=blue, citecolor=blue]{hyperref}

\begin{document}

\author{Ran Wei}
\affiliation{Hefei National Laboratory for Physical Sciences at Microscale and Department
of Modern Physics, University of Science and Technology of China, Hefei, Anhui
230026, China}
\affiliation{Laboratory of Atomic and Solid State Physics, Cornell University}
\author{Erich Mueller}
\affiliation{Laboratory of Atomic and Solid State Physics, Cornell University}
\title{Vortex structures of a two-component BEC for large anisotropies}
\date{\today}

\begin{abstract}
We calculate the vortex structures of an elongated two-component Bose-Einstein condensate.
We study how these structures depend on the intra-component and inter-component interaction strengths.  We present analytic and numeric results respectively at weak and strong interactions; finding lattices with different interlocking geometries: triangular, square, rectangular and double-core.
\end{abstract}

\maketitle

\section{Introduction}

One of the most exciting recent developments in cold atom experiments has been the production of artificial gauge fields, which couple to neutral atoms in the same way that magnetic fields couple to charged particles \cite{Lin2009nature,Spielman2009}. While the greatest excitement surrounds the possibility of producing analogs of fractional quantum Hall states or topological insulators \cite{spielmantheory},
these experiments also allow one to perform some extremely interesting experiments on vortices in Bose-Einstein condensates.  These structures are particularly rich for multi-component gases \cite{Mueller2002,Ueda2003,Kasamatsu2009,Wu2011}. A key difference between the latest NIST experiments, and previous ones on rotating condensates \cite{Cornell1999,Dalibard2000,Ketterle2001,Cornell2001,Cornell2004,Ketterle2005},
is that the techniques used to generate the artificial gauge fields naturally lead one to consider very narrow geometries. Here we theoretically investigate the vortex structures in a two-component Bose-Einstein condensate confined to a relatively narrow channel.  We find a rich phase diagram which complements our understanding of the isotropic 2D system \cite{Mueller2002,Ueda2003,Kasamatsu2009,Wu2011}.

In the NIST experiments the hyperfine states of $^{87}$Rb are coupled by a series of Raman lasers.  In the dressed state picture \cite{Lin2009nature,Spielman2009}, taking the strong Raman field limit, one arrives at an effective single particle Hamiltonian
\begin{equation}\label{sp}
H_1=\left[p_x-A_x(y)\right]^2/2m + p_y^2/2m + p_z^2/2m + V_{\rm eff}.
\end{equation}
The effective gauge field is bounded, $-\hbar k_R<A_x(y)<\hbar k_R$,  where $\hbar k_R$ is the momentum kick from absorbing a photon from one beam and releasing it into another.  One can generate arbitrarily large magnetic fields $B=\partial_y A_x$, but due to the bounded nature of $A_x$, these can only exist over a finite extent in the $y$ direction.  It is therefore natural to choose the effective potential $V_{\rm eff}(y,z)$ to restrict the particles to a small region of $y,z$, modeling it as $V_{\rm eff}(y,z)=m{\omega_y^*}^2 y^2/2+m{\omega_z}^2 z^2/2$.  $V_{\rm eff}$ includes both the external potential and one induced by the Raman lasers, and $\omega_{y}^*$ is an effective trap frequency.

The impact of the magnetic field on a single component Bose-Einstein condensate in such a narrow channel has been extensively studied \cite{Sinha2005,Palacios2005,Shlyapnikov2009}. As one increases the magnetic field from zero, the Bogoliubov spectrum starts to develop a ``roton" minimum, whose energy falls with increasing $B$.  When the roton energy hits zero, the system becomes unstable to undulations.  At higher magnetic fields vortices enter the channel.  Here we extend these results to the two-component gas.

The vortex structures in isotropic 2D two-component condensates are quite rich \cite{Ueda2003,Mueller2002}.  Depending on the relative strength of the inter-component and intra-component scattering one can find interpenetrating lattices with different geometries: triangular, square, rectangular and double-core.  Additionally, one sees interesting hysteresis \cite{Ueda2003}.  We find similar results in the anisotropic geometry.

A final motivation for thinking about the role of magnetic fields in very narrow geometries comes from solid state physics.  In 2005, Seidel et al \cite{Seidel2005} showed that the quantum Hall effect in narrow tori can be connected to charge density waves, with the fractionally charged vortices mapping onto fractionally charged domain walls.  In this paper we will focus on the Bose condensed regime, and will not be able to comment on this interesting physics.  We do note, however, that our system undergoes a charge density wave instability before vortices enter the system.  Investigating this instability in the low density ``quantum" limit may be fruitful, even though the physics is likely to be very different from that in \cite{Seidel2005}.

\section{model}

There are four important lengths in Eq.~(\ref{sp}), $d_z=\sqrt{\hbar/m\omega_z}, d_y=\sqrt{\hbar/m\omega_y^*},
\ell=\sqrt{\hbar/\bar B}$, and $R=\hbar k_R/\bar B$, where $\bar B$ is the peak value of $B$.  These correspond to the trap length in the $z$ and $y$ direction, the magnetic length, and the spatial range over which the effective magnetic is nonzero.
If $d_z$ is sufficiently small, the extent in $z$ direction is sufficiently compressed and the 3D gases can be treated as 2D cloud. As we argue below, a sufficient condition will be $d_z\lesssim\ell$.
If $d_y\ll R$, the cloud will be confined to a region where the vector potential varies linearly and we can
approximate: $A_x=\bar B y$.  In terms of the raw experimental parameters, $\bar B=\delta'\hbar k_R/\Omega$, where $\Omega$ is the Raman Rabi frequency, and $\delta'$ is the gradient of two-photon detuning along $y$ direction.

Taking $\Psi_1({\bf r})$ and $\Psi_2({\bf r})$ to annihilate atoms in the two (pseudo)-spin states, the short range interactions will be
\begin{eqnarray}
H_{\rm int} &=&\frac{1}{2}\int d{\bf r}(g_1\Psi^{\dagger}_{1}\Psi^{\dagger}_{1}\Psi_{1}\Psi_{1}+
g_2\Psi^{\dagger}_{2}\Psi^{\dagger}_{2}\Psi_{2}\Psi_{2}\nonumber\\&&\qquad\qquad\qquad+
2g_{12}\Psi^{\dagger}_{1}\Psi^{\dagger}_{2}\Psi_{2}\Psi_{1}).
\end{eqnarray}
The coupling constants are related to scattering lengths $a_s$ via $g=4\pi\hbar^2 a_s/m$.  We will consider the case $g_1=g_2=g$, and $g_{12}=g_m$. For most experiments $g_1\approx g_2\approx g_{12}$, but one gains insight by relaxing this condition.

Following Sinha and Shlyapnikov \cite{Sinha2005}, we diagonalize the single particle Hamiltonian in Eq.~(\ref{sp}).  The eigenstates are labeled by three quantum numbers $K,n,n^\prime$, with energies
\begin{eqnarray}
E_{nn^\prime}(K)&=& {\cal E} K^2 + n \hbar\omega_z+n^\prime \hbar \tilde \omega_c.
\end{eqnarray}
where  $\tilde \omega_c^2= {\omega_y^*}^2+\omega_c^2$, $\omega_c=\bar B/m$,
and ${\cal E}=\hbar^2{\omega_y^*}^2/4m\tilde\omega_c^2\tilde\ell^2$,
$\tilde\ell=\sqrt{\hbar/m\tilde\omega_c}$, and we have neglected the zero-point energy.
The continuous variable $K=\sqrt{2}\tilde\ell k$ is a dimensionless label of the momentum $k$ along the $x$ direction, while $n$ and $n^\prime$ are discrete quantum numbers corresponding to the number of nodes in the $z$ and $y$ directions.
If the interaction energy per particle $\langle H_{\rm int}/N\rangle$ and the characteristic ``kinetic energy" $\langle {\cal E} K^2\rangle$ are small compared to $\hbar \tilde\omega_c$ and $\hbar\omega_z$
one can truncate to the single particle
eigenstates with $n=n^\prime=0$, which are of the form
\begin{equation}\label{sp1}
\phi_K(x,y,z)=\frac{1}{\sqrt{\pi\tilde\ell d_zL}}e^{i\frac{Kx}{\sqrt{2}\tilde\ell}}e^{-\frac{(y-y_K)^2}{2\tilde\ell^2}}e^{-\frac{z^2}{2d_z^2}}
\end{equation}
where $y_K=\sqrt{2}\omega_cK\tilde\ell/2\tilde\omega_c$ and
$L$ is the length of $x$ direction.

By taking the system sufficiently dilute, it is easy to arrange  $\langle H_{\rm int}/N\rangle\ll \hbar\tilde\omega_c\sim\hbar\omega_z$.  The other condition,
${\cal E} K^2 \ll \hbar \tilde \omega_c$, will be valid at strong magnetic fields.  For our vortex lattices, we find the characteristic dimensionless wave-number to be $K\sim1$, thus requiring $\omega_y^*\ll\omega_c$.
Combining this with the previous constraint, $\omega_y^*\gg m\omega_c^2/\hbar k_R^2$,
we see that our approximations break down unless the magnetic length is much larger than the wavelength of the Raman lasers, ie. $\omega_c\ll\hbar k_R^2/m$. This also establishes our requirement $d_z\lesssim\ell$.

Letting $a_K$ annihilate the state in Eq.~(\ref{sp1}), the $N$-body Hamiltonian is
\begin{eqnarray}
H/{\cal E}&=&\sum_KK^2\left[a_{K\uparrow}^\dagger a_{K\uparrow}+ a_{K\downarrow}^\dagger a_{K\downarrow}\right]
+\frac{\beta}{N}\sum_{q}F^\dagger_{\uparrow}(q)F_{\uparrow}(q)\nonumber\\
&+&\frac{\beta}{N}\sum_{q}F^\dagger_{\downarrow}(q)F_{\downarrow}(q)
+\frac{2\beta_m}{N}\sum_{q}F^\dagger_{\uparrow\downarrow}(q)F_{\uparrow\downarrow}(q)
\end{eqnarray}
where
\begin{eqnarray}
F_{\sigma\tau}(q)=\sum_{K_1K_2}\delta_{q-K_1-K_2}e^{-\frac{1}{8}(K_1-K_2)^2}a_{K_1\sigma}a_{K_2\tau}.
\end{eqnarray}
We take the continuum limit $\sum_K\to (\sqrt{2}L/4\pi\tilde\ell)\int dK$, and $\delta_K\to (2\sqrt{2}\pi\tilde\ell/L) \delta(K)$.  The effective 1D parameters are $\beta=Nmg\tilde\omega_c^2\tilde\ell/\pi Ld_z{\omega_y^*}^2\hbar^2, \beta_m=\beta g_m/g$. From these definitions of the effective interaction
parameters $\beta,\beta_m$, we see that increasing $g,g_m$ has the same effect as increasing the magnetic field $B$, increasing the confinement $\omega_z$, or reducing the confinement $\omega_y^*$.

We consider a variational wavefunction corresponding to a Bose-Einstein condensate which is periodic in the $x$ direction, with dimensionless wavelength $2\pi/K_0$, corresponding to a physical wavelength $\lambda=2\sqrt{2}\pi\tilde\ell/K_0$
\begin{equation}\label{coher}
|\psi\rangle =\exp \left[-\frac{N}{2}+\sum_{\sigma} \sqrt{N_\sigma}\left( \sum_n C_{n\sigma} a_{n K_0 \,\sigma}^\dagger \right) \right]|\mbox{vac}\rangle.
\end{equation}
Where $\sigma=\uparrow,\downarrow$ for two-component BEC.
$N_\sigma$ is the number of particles in state $\sigma$, and $N=N_\uparrow+N_\downarrow$.
The coefficients $C_{n\sigma}$ are normalized to $\sum_n |C_{n\sigma}|^2=1$.
In place of such a coherent state ansatz, some authors prefer to work with a ``Fock" state
\begin{eqnarray}\label{fock}
\notag &  &|\psi\rangle _F = \frac{1}{\sqrt{N_\uparrow! N_\downarrow!}}\\
&  &\left( \sum_n C_{n\uparrow} a_{n K_0 \,\uparrow}^\dagger \right)^{N_\uparrow}
\left( \sum_n C_{n\downarrow} a_{n K_0 \,\downarrow}^\dagger \right)^{N_\downarrow} |\mbox{vac}\rangle.
\end{eqnarray}
For the quantities we are interested in, $|\psi\rangle$ and $|\psi\rangle_F$ are equivalent, and the variational parameters $C_{n\sigma}$ have the same meaning in each case: $|C_{n\sigma}|^2$ is the fraction of spin $\sigma$ particles with momentum $n K_0$.

We take $N_\uparrow=N_\downarrow$, and further restrict ourselves
to considering symmetric or antisymmetric wavefunctions:
$C_{n,\sigma}=\pm C_{-n,\sigma}$.  We classify our state by the value of $K_0$ and $\xi$ the number of non-zero $C_n$'s which are needed to minimize the energy.  For example, as illustrated in Fig.~\ref{pd}, for $(\beta,\beta_m)\sim(3,1)$,
we need only one $n$ for each component: $n=0$.  We refer to this state as $(\xi_\uparrow,\xi_\downarrow)=(1,1)$.
This should be contrasted with the case at  $(\beta,\beta_m)\sim(5,2.5)$, where the energy is minimized by taking $n=-1,0,1$ in both components, a state we label as $(\xi_\uparrow,\xi_\downarrow)=(3,3)$.  Analytic expressions for the energies with small numbers of components are given in appendix~\ref{aplab}.

\section{Results at small $\beta,\beta_m$}

\begin{figure}[!htb]
\includegraphics[width=7cm]{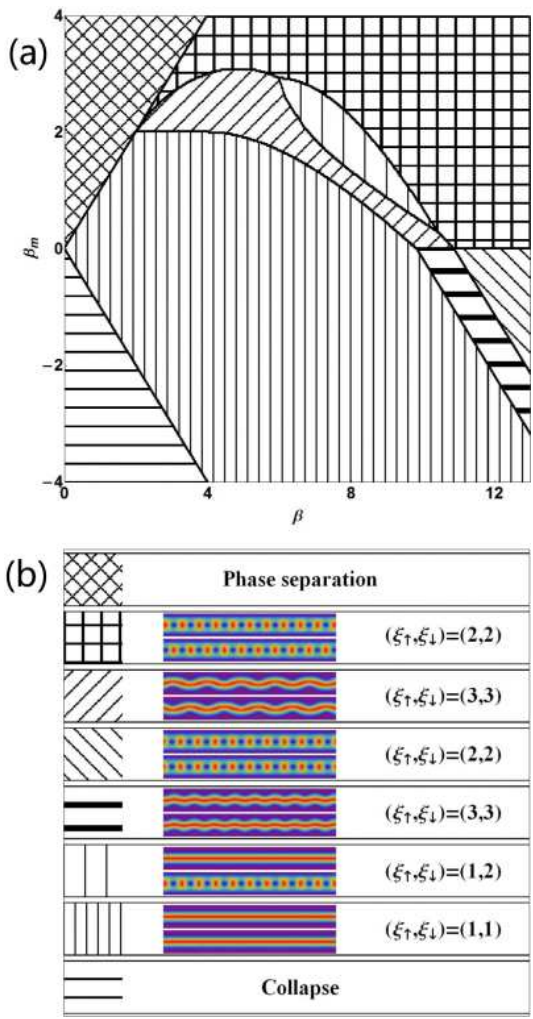}
\caption{Top panel: phase diagram at small dimensionless intra-species and inter-species interactions
$\beta,\beta_m$. Larger $\beta,\beta_m$ corresponds to larger $g,g_m$, larger
magnetic field, larger confinement $\omega_z$, or weaker confinement $\omega_y^*$.
Hatched patterns represent states described by different number of Fourier components in each spin state:
($\xi_\uparrow,\xi_\downarrow$).
Bottom panel: the density profiles of the two-component wavefunction in the corresponding regimes.
A color key for the density patterns is shown in Fig.\ref{lattices}.}
\label{pd}
\end{figure}

The number of expansion parameters, $(\xi_\uparrow,\xi_\downarrow),$ grow with the magnitude of $\beta,\beta_m$.
For small $\beta$ and $\beta_m$, we only need a small number of terms in our wavefunctions, and we may use the analytic expressions in Eq.~\ref{ap1}-\ref{ap2} to find the lowest energy state. We obtain a series of phase, as demonstrated in Fig.\ref{pd}(a).

\begin{figure}[!htb]
\includegraphics[width=8cm]{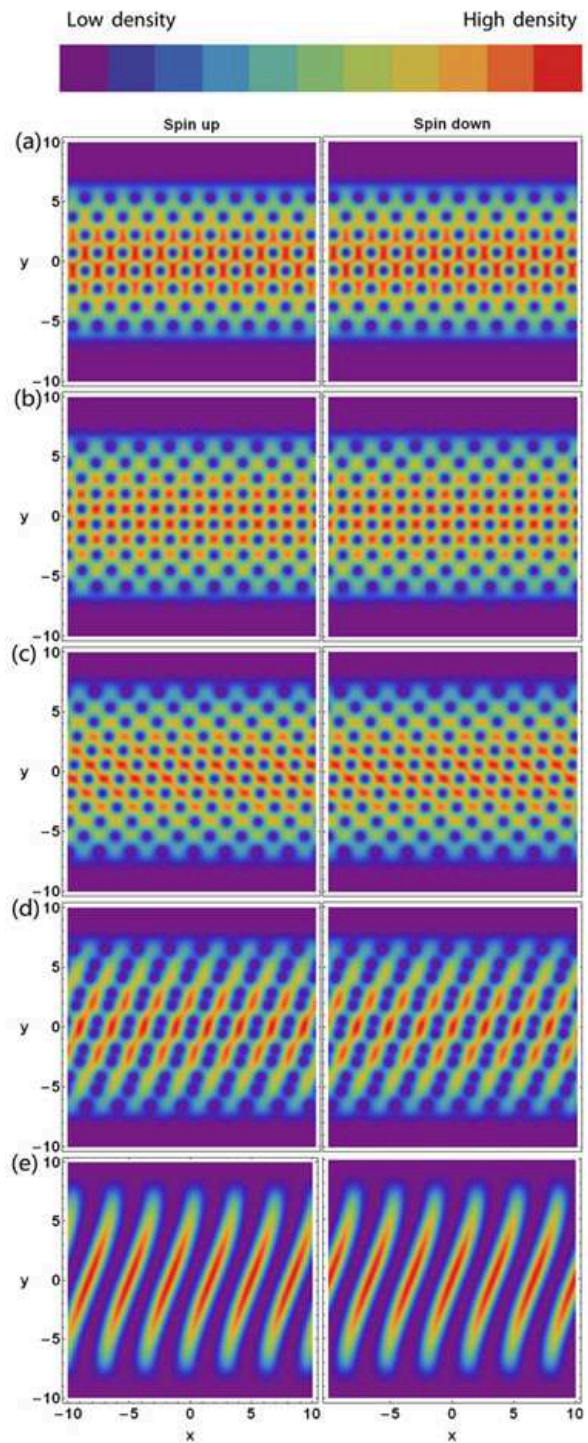}
\caption{The density profiles of two-component BEC at large $\beta,\beta_m$,
where $x,y$ are in the unit of $\sqrt{2}\tilde\ell$.
For $(a)-(e)$, the parameters are
$\beta_m/\beta=0.1,0.5,0.8,1,1.5$ respectively, with $\beta=1000$.
A color key is shown at the top.}
\label{lattices}
\end{figure}

At $\beta_m=0$, the $\uparrow$ and $\downarrow$ atoms decouple, and the physics is identical to the single component case, $\xi_\uparrow=\xi_\downarrow=\xi$. When $\beta<9.8$, the ground state wavefunction has only one term ($\xi=1$), which displays a Gaussian shape along $y$ direction. As $\beta$ increases past $\beta=9.8$, the $\xi=1$ wavefunction becomes unstable, and it undergoes a second-order phase transition to a density wave with $\xi=3$. As $\beta$ increases to $\beta=10.8$, a first-order phase transition occurs to a state with $\xi=2$, characterized by a single row of vortices. These results have been extensively studied in Ref.~\cite{Shlyapnikov2009}.

In the regime of repulsive inter-component interaction ($\beta_m>0$), the $\uparrow$ and $\downarrow$ particles try to avoid each-other. For strong repulsive interaction ($\beta_m>\beta$), the two components undergoes a microscale phase separate~\cite{Timmermans1998}, which needs large (presumably infinite) $xi$ to describe. For weak repulsive interaction ($\beta_m<\beta$) rich structures, illustrated in Fig.~\ref{pd} develop.
For some $\beta,\beta_m$,
we find $\xi_\uparrow\neq\xi_\downarrow$.  For example, at $(\beta,\beta_m)=(8,2)$, $\xi_\uparrow=1$ and $\xi_\downarrow=2$, and the $\downarrow$ atoms have a row of
vortices, while the $\uparrow$ atoms show no structure.  Under these circumstances there is a degenerate state with the $\uparrow$ and $\downarrow$ wavefunctions reversed.
For other $\beta,\beta_m$, there is a symmetry between the two components with $\xi_\uparrow=\xi_\downarrow$.  The vortices or density corrugations are displaced by half a period so that the density maxima of the $\uparrow$ atoms line up with the density minima of the others.
For example, when $(\beta,\beta_m)=(10,3)$, each component displays a single row vortices, and the wavefunctions are related by a translation.

In the regime of attractive inter-component interaction ($\beta_m<0$), one wants to maximize the overlap of  the $\uparrow$ and $\downarrow$ wavefunctions.  Generically, this means that the wavefunction of each component is identical, and the problem reduces to the single component case, but with a renormalized interaction $\beta\to\beta+\beta_m$.  The
phase diagram for $\beta_m<0$ can be calculated from the phase diagram at $\beta_m=0$ by mapping each point: $(\beta,\beta_m)\rightarrow(\beta-\beta_m,0)$.
For strong attractive interaction ($-\beta_m>\beta$), the BEC is unstable and expected to
collapse, similar to the case of a single component BEC with attractive interaction~\cite{Hulet2000}.

\section{Results at large $\beta,\beta_m$}

For large $\beta$ and $\beta_m$, the analytic expressions for the energy become unwieldy.
We numerically minimize the expectation $\langle H\rangle$, varying $\{ C_{n\sigma},K_0\}$ in
Eq.~(\ref{coher}).  The energy landscape has many local minima, and we use a range of starting parameters to try to find the
absolute minimum.  We cannot rule out the existence of even lower energy states.  Moreover, for some parameters we found that the
energy differences between competing minima became extremely small.  In an experiment it is doubtful that one would find
the true minimum energy state.  Rather than systematically exploring the large $\beta$ physics, we simply show a few examples.

Our results, illustrated in Fig.~\ref{lattices}, are richer than those seen in the single component gas~\cite{Shlyapnikov2009},
showing structures similar to those in isotropic 2D studies \cite{Mueller2002,Ueda2003}.
When $\beta_m\ll\beta$ one finds two interlocking triangular lattices, as in Fig.~\ref{lattices}(a) where $\beta_m/\beta=0.1$.
As one increases $\beta_m$ the lattice structure changes: interlocking square/rectangular lattices are shown in Fig.~\ref{lattices}(b,c) where $\beta_m/\beta=0.5,0.8$.
When $\beta_m>\beta$ the non-rotating system would be expected to phase separate.  
Here, one finds more intricate vortex structures at $\beta_m\sim\beta$.
At $\beta_m/\beta=1$, we find double-core vortices (cf. \cite{Ueda2003}), as in Fig.~\ref{lattices}(d).  At $\beta_m/\beta=1.5$ we find stripes, which are a microscopic version of phase separation.  Close inspection of the image in Fig.~\ref{lattices}(e), shows vortex cores in the low density regions.

\section{Summary and Conclusions}

We have investigated the vortex structures in a two-component BEC with an artificial magnetic field, in an elongated geometry. Compared to the single component gas, the two-component vortex structures are more intricate.

To experimentally investigate these structures, one needs to find a system where the interspecies interactions can be tuned relative to the intraspecies.  One promising approach is to use different atomic species for the two (pseudo)-spin states, and take advantage of an interspecies Feshbach resonance \cite{interspeciesfeshbach}. Some of the other newly condensed atomic systems may also be favorable \cite{novel}.  If one cannot separately tune $\beta$ and $\beta_m$, one can still change the magnitude of them, fixing $\beta_m/\beta$.  As seen in Fig.~\ref{pd}, such a cut through the phase diagram can still be quite rich, especially if $\beta_m/\beta\sim0.25$.

\section{Acknowledgment}
R. W. is supported by CSC, the NNSFC, the NNSFC of Anhui (under Grant No. 090416224), the CAS, and the
National Fundamental Research Program (under Grant No. 2011CB921304). This material is based upon work supported by the National Science Foundation under Grant No. PHY-1068165.

\appendix
\section{Analytic results for small number of components.}\label{aplab}
Here we give analytic results for the energies of the states defined in Eq.~(\ref{coher}), truncating the $n$-sums for each $\sigma$, and assuming symmetry/antisymmetry about the origin for each component.

State $(\xi_\uparrow,\xi_\downarrow)=(1,1)$ is unique, and has energy
\begin{eqnarray}
\label{ap1}
E_{1,1}=\frac{1}{2}(\beta+\beta_m)
\end{eqnarray}

State $(\xi_\uparrow,\xi_\downarrow)=(1,2)$, with
\begin{eqnarray}
\psi_{\uparrow}&=&\phi_{0\uparrow}\\
\psi_{\downarrow}&=&\frac{\sqrt{2}}{2}(\phi_{K_0\downarrow}+\phi_{-K_0\downarrow})
\end{eqnarray}
is unique up to translation, and has energy
\begin{eqnarray}
E_{1,2}=\frac{1}{2}K_0^2+\frac{1}{4}\beta(e^{-K_0^2}+\frac{3}{2})+\frac{1}{2}\beta_me^{-\frac{1}{4}K_0^2}
\end{eqnarray}

State $(\xi_\uparrow,\xi_\downarrow)=(1,3)$, with
\begin{eqnarray}
\psi_{\uparrow}&=&\phi_{0\uparrow}\\
\psi_{\downarrow}&=&\sqrt{1-2|\varepsilon|^2}\phi_{0\downarrow}-\varepsilon(\phi_{K_0\downarrow}+\phi_{-K_0\downarrow})
\end{eqnarray}
has energy
\begin{equation}
\begin{split}
E_{1,3}&=\frac{1}{2}(\beta+\beta_m)\\+&\left[K_0^2+\beta(2e^{-\frac{1}{4}K_0^2}-e^{-\frac{1}{2}K_0^2}-1)
+\beta_m(e^{-\frac{1}{4}K_0^2}-1)\right]\epsilon^2\\
+&\beta\left(e^{-K_0^2}+2e^{-\frac{1}{2}K_0^2}-4e^{-\frac{1}{4}K_0^2}+\frac{3}{2}\right)\epsilon^4
\end{split}
\end{equation}

For state $(\xi_\uparrow,\xi_\downarrow)=(2,2)$, there are two distinct extremal states.  Both components could be symmetric about the origin ($C_{1\uparrow}=C_{-1\uparrow}=C_{1\downarrow}=C_{-1\downarrow}=1/\sqrt{2}$), or one of them could be antisymmetric ($C_{1\uparrow}=C_{-1\uparrow}=C_{1\downarrow}=-C_{-1\downarrow}=1/\sqrt{2}$). This gives energies
\begin{eqnarray}
E^s_{2,2}&=&K_0^2+\frac{1}{4}(\beta+\beta_m)+\frac{1}{2}(\beta+\beta_m)e^{-K_0^2}\\
E^a_{2,2}&=&E^s_{2,2}-\frac{1}{2}\beta_me^{-K_0^2}
\end{eqnarray}

For state $(\xi_\uparrow,\xi_\downarrow)=(2,3)$, the $\uparrow$ component can be symmetric or antisymmetric, resulting in energies
\begin{eqnarray}
\notag E^s_{2,3}&=&\frac{1}{2}K_0^2+\frac{1}{4}\beta(e^{-K_0^2}+\frac{3}{2})+\frac{1}{2}\beta_me^{-\frac{1}{4}K_0^2}\\\nonumber
&+&\left[K_0^2+\beta(2e^{-\frac{1}{4}K_0^2}-e^{-\frac{1}{2}K_0^2}-1)\right]\epsilon^2\\
&+&\frac{1}{2}\beta_m(1-2e^{-\frac{1}{4}K_0^2}+2e^{-K_0^2})\epsilon^2\\\nonumber
&+&\beta(\frac{3}{2}-4e^{-\frac{1}{4}K_0^2}+2e^{-\frac{1}{2}K_0^2}+e^{-K_0^2})\epsilon^4\\
E^a_{2,3}&=&E^s_{2,3}-\beta_m e^{-K_0^2} \epsilon^2
\end{eqnarray}

For state $(\xi_\uparrow,\xi_\downarrow)=(3,3)$, the optimal wavefunction is
\begin{eqnarray}
\psi_{\uparrow}=\sqrt{1-2|\epsilon|^2}\phi_{0\uparrow}+i \epsilon(\phi_{K_0\uparrow}+\phi_{-K_0\uparrow})\\
\psi_{\downarrow}=\sqrt{1-2|\epsilon|^2}\phi_{0\downarrow}\pm i \epsilon(\phi_{K_0\downarrow}+\phi_{-K_0\downarrow})
\end{eqnarray}
with $\epsilon$ real.

If $\beta>\beta_m>0$, the negative sign has lower energy,
\begin{eqnarray}
\nonumber E^-_{3,3}&=&\frac{1}{2}(\beta+\beta_m)\\\nonumber
&+&\left[2K_0^2+\beta(-2e^{-\frac{1}{2}K_0^2}+4e^{-\frac{1}{4}K_0^2}-2)\right]\epsilon^2\\\nonumber
&+&\beta_m(2e^{-\frac{1}{2}K_0^2}-2)\epsilon^2\\
&+&\beta(2e^{-K_0^2}+4e^{-\frac{1}{2}K_0^2}-8e^{-\frac{1}{4}K_0^2}+3)\epsilon^4\\\nonumber
&+&\beta_m(2e^{-K_0^2}-4e^{-\frac{1}{2}K_0^2}+3)\epsilon^4
\end{eqnarray}
If $\beta>0,\beta_m<0$, the positive sign has lower energy,
\begin{eqnarray}
\label{ap2}
E^+_{3,3}=E^-_{3,3}+4 \beta_m (e^{-K_0^2/4}-e^{-K_0^2/2}) (\epsilon^2+2 \epsilon^4)
\end{eqnarray}

\end{document}